# Superradiant Electron Energy Loss Spectroscopy

Ron Ruimy[†], Alexey Gorlach[†], Gefen Baranes Spitzer, and Ido Kaminer

† equal contributors

*Solid State Institute and Faculty of Electrical & Computer Engineering, Technion-Israel Institute of Technology, Haifa 32000, Israel*

kaminer@technion.ac.il

We analyze the interaction between a free electron and an ensemble of identical optical emitters. The mutual coherence and correlations between the emitters can enhance the interaction with each electron and become imprinted on its energy spectrum. We present schemes by which such collective interactions can be realized. As a possible application, we investigate free-electron interactions with superradiant systems, showing how electrons can probe the ultrafast population dynamics of superradiance.

# 1. Introduction

Electron microscopy and spectroscopy are powerful analytical tools for extracting information about quantum emitters such as atoms, molecules, vacancy centers, and quantum dots [1]. Specifically, electron-based approaches such as cathodoluminescence [1] and electron energy-loss spectroscopy (EELS) [1] can probe excitation energies, atomic bandgaps, local density of photonic states, and recently even observe phononic-phenomena such as vibrational modes in molecules [2]. In all these examples, free-electron probe techniques can provide very high spatial and energy resolutions.

An emerging field of electron microscopy and spectroscopy is *imaging quantum coherence* of bound electron systems [3-7] and even of light [8,9,10]. Much theoretical effort in recent years [3-5,11] investigated the coherent interaction between free electrons and discrete quantum systems with energy gap and transitions in the optical range. Such interactions have been proposed as a mean to measure strong-coupling physics with deep-subwavelength resolution [12], extract the coherent quantum states of quantum emitters [4,5], generate entanglement between them [5], and even control their quantum state [3]. A key element necessary for these capabilities is the coherent shaping of the free-electron wavefunction in the time domain, through photon-induced nearfield electron microscopy (PINEM) [13-23]. Such shaped electrons can undergo a coherent resonant interaction with a quantum emitter, in a process also called free-electron-bound-electron resonant interaction (FEBERI) [3] or quantum klystron in microwave frequencies [24]. So far, due to the intrinsically weak coupling between free electrons and bound electrons [4,11], experimental realizations of these phenomena have remained beyond reach.

High densities of emitters hold importance for applications [25-27] such as semiconductor quantum dot devices for quantum science and technology [28-30]. High emitter densities are a fundamental aspect of certain light-matter interactions such as superradiance [31-34] and superfluorescence [35,36]. Similarly, the existence of multiple emitters interacting with each other is fundamental for quantum information and computation problems [37], especially ones that involve collective phenomena and correlations between emitters. These kinds of correlations typically occur at the nanometer scale, making it particularly challenging to probe them with optical techniques, due to the diffraction limit. Those considerations motivate our proposal to use free electrons to probe high-density ensembles of emitters, toward the investigation of many-body quantum physics using free-electron probes. However, so far, the coherent interaction between free electrons and ensembles of quantum systems has yet to be described

Here we analyze the coherent interaction between free electrons and an ensemble of identical optical emitters (Fig. 1). The electron interaction can be sensitive to the many-body nature of such an ensemble of emitters, strongly enhanced by multi-emitter correlations and mutual coherence. We examine these enhanced interactions for the exemplary case of superradiant dynamic [31,32], extracting the entire population statistics of the superradiant states using the electrons. These prospects rely on a key realization: that the multi-emitter correlations that enable superradiance also *enhance their interaction with each free electron*. Our work contributes to currents efforts toward the investigation of general many-body quantum phenomena using electrons, aiming to achieve simultaneous sub-nanometer and femtosecond resolution using experimental setups such as ultrafast transmission electron microscopy (UTEM) (Fig. 1d) [12,18-20,38].

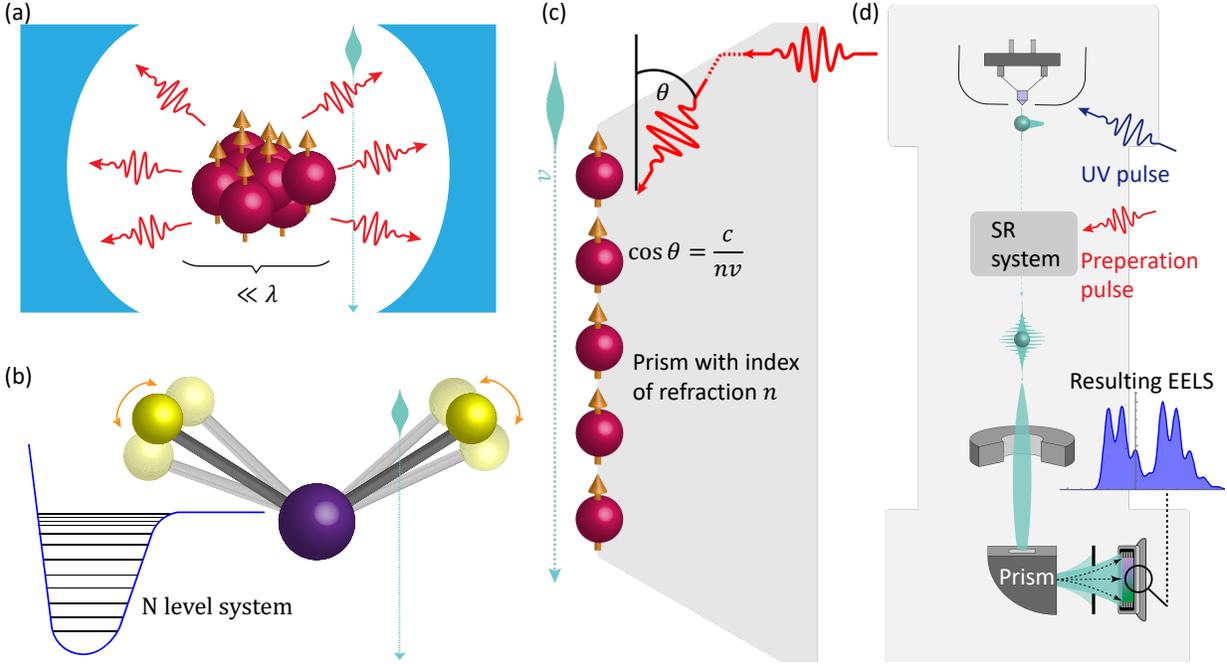

**Fig. 1. Superradiant electron energy loss spectroscopy (EELS)**: systems containing multi-emitter correlations that can be investigated using EELS. **(a)** High density of emitters confined in a small volume compared to the emitted light wavelength can interact superradiantly. **(b)** Vibrational modes in molecules are often described by anharmonic oscillators, creating effective N-level systems that mimic superradiant phenomena. **(c)** Identical quantum emitters arranged along a surface can be synchronized to behave superradiantly even when they are relatively far apart, using their joint interaction with a single electron that correlates between them. We propose such a scenario in a grazing-angle configuration [21] and find that it is governed by a Cherenkov-type phase-matching condition. **(d)** Superradiant EELS can be realized in an ultrafast transmission electron microscope. A pump laser pulse prepares the state of the initial superradiant system, which then interacts with the electron at a delay depicted by a probe pulse. We can extract the state of the superradiant system and its ultrafast dynamics from the EELS measurement.

## 2. Theory of electron interactions with ensembles of emitters

We model each emitter as a general multi-level system that is small enough to be treated in the dipole approximation. The emitters are coupled to the near-field of the free electron through dipole interaction. We calculate the electron–emitters interaction scattering matrix. The $i$'th emitter is located at position $z_i$ with impact parameter $r_\perp^i$. Each emitter's energy states can be denoted as $|n^i\rangle$, and the energy separation between the states is denoted as $\hbar\omega_i^{nm}$. Assuming paraxiality and weak interaction strength with each individual emitter, the coupling strength between an electron and an emitter is (see [4] for a discussion on these conditions and their relevance for experiments):

$$g_i^{nm} = \left( \frac{ed_{\perp,nm}^i \omega_i^{nm} K_1\left(\frac{\omega_i^{nm} r_\perp^i}{\gamma v}\right)}{2\pi\varepsilon_0 \hbar v^2} + \frac{ed_{z,nm}^i \omega_i^{nm} K_0\left(\frac{\omega_i^{nm} r_\perp^i}{\gamma v}\right)}{2\pi\gamma\varepsilon_0 \hbar v^2} \right) e^{-\frac{i\omega_i^{nm} z_i}{v}}. \quad (1)$$

Here $v$ is the average velocity of the electron wavepacket; $\boldsymbol{d}_{nm}^i = \langle n^i|e\boldsymbol{r}|m^i\rangle$; the vacuum permittivity is $\varepsilon_0$; the Lorentz factor is $\gamma$; and $K_0(x), K_1(x)$ are modified Bessel functions of the second kind.

The scattering matrix is given by:

$$\mathbb{S} = e^{i \sum_{i,n,m} g_i^{nm} t_i^{nm} b_i^{nm}}, \quad (2)$$

where the operators $b_i^{nm}$ are the momentum translation operators [39] with translation of $-\hbar\omega_i^{nm}/v$. The operators $t_i^{nm}$ are the emitters' transition operators $|n_i\rangle\langle m_i|$. The Hilbert space of the emitters is hard to analyze because of its size, which is generally $K^N \times K^N$, where $N$ is the number of emitters and $K$ is the number of energy levels in each emitter.

Eq. (2) is completely general, however for the purpose of this manuscript we are interested in limited case that capture some of the key phenomena of electron interactions. We assume that all the emitters are identical in terms of transition dipoles and energy levels, differing only in their location on the z axis, creating an effective 1D structure (Fig. 1c). Furthermore, we ignore all energy states except the ground state and some excited state with energy separation $\hbar\omega_0$, making the emitters effective two-level systems. This assumption is justified typically when there is one transition dipole moment more dominant than the rest, and when the excitation of the emitter is done with a narrow-band pulse making the population of the other energetic states initially negligible. Furthermore, since this manuscript focuses on resonances arising from the phase matching between an electron and a specific transition, this transition becomes dominant compared to the other transitions (especially in the case of many emitters). In this case, the scattering matrix can be simplified to

$$\mathbb{S} = e^{i(gS_+ b + g^* S_- b^\dagger)}. \tag{3}$$

$g$ is the coupling strength defined in Eq. (1) and $S_\pm = \sum_i \sigma_\pm^i e^{\mp \frac{i\omega_0 z_i}{v}}$ where $\sigma_\pm$ are Pauli matrices. These super operators define a superradiant ladder that can be constructed by acting with these operators on the ground state $|m\rangle = \frac{1}{\sqrt{\binom{N}{m}}} S_+^m |gg \dots g\rangle$. The scattering matrix is closed in the sub-Hilbert space of this superradiant ladder. If the initial state of the emitters is in this sub-Hilbert space, then the scattering matrix is reduced to be $(N+1) \times (N+1)$ for the emitters. A detailed derivation is presented in SM section 1.

To understand the in what sense this ladder is superradiant, we write the $S_\pm$ operators in the interaction picture (with respect to the emitters free Hamiltonian):

$$S_\pm^I = \sum_i \sigma_\pm^i e^{\pm i\omega_0 \left(t - \frac{z_i}{v}\right)}. \tag{4}$$

The electron arrives to the emitter located at $z_i$ at time $z_i/v$ and thus experiences a delayed interaction with the different emitters. From the reference frame of the electron, these operators are Dicke superradiance ladder operators of a sub-Hilbert space of symmetric states that are phase shifted according to the corresponding delays. If the emitters are initially at the ground state, the electron excitation keeps them inside this symmetric (superradiant) sub-Hilbert space spanned by the operators $S_\pm$. Also, if the emitters are pre-excited to satisfy phase-matching condition with the electron, the interaction keeps them inside this superradiant sub-Hilbert space. From the perspective of the electron, all the transition dipoles add up constructively as is the typical case in Dicke superradiance [31], resulting in an effective dipole with size $N \cdot d$. Note that the operators in Eq. (4) reproduce conventional Dicke superradiance [31,32] when the emitters are located close enough to each other, typically within a single wavelength [32] (Fig. 1a). The same mathematical framework we develop also applies to equally spaced $N$-level systems such as the anharmonic oscillator modeling of vibrational states in molecules [40] (Fig. 1b).

The elements of the scattering matrix can be found analytically and are given by:

$$s_{nm} = \langle n|S|m \rangle = b^{n-m}(\cos|g|)^N (i\tan|g|)^{n-m} \times$$

$$\sqrt{m!\,n!\,(N-n)!\,(N-m)!} \sum_{k=0}^{m} \frac{(-1)^k (\tan|g|)^{2k}}{k!\,(m-k)!\,(n-m+k)!\,(N-n-k)!}. \tag{5}$$

If we assume that $|g| \ll 1$ and additionally $N - m \gg 1$ and $m \gg 1$, meaning that the state is far from the edges of the ladder, then Eq. (5) can be further simplified (see SM section 2):

$$s_{nm} \approx J_{n-m}\left(2|g|\sqrt{Nm - m^2}\right) e^{i(n-m)\arg(g)}. \tag{6}$$

In this approximation, the scattering matrix approaches that of a coherent interaction between a free electron and light, as in photon-induced nearfield electron microscopy [16]. The similarity arises when the photonic excitation in its quantum harmonic oscillator is far from the ground state, and then it resembles the ladder states far from the edges. The resulting effective coupling equals $g_{\text{eff}} = |g|\sqrt{Nm - m^2}$. If we imagine for example that the emitters are excited using a coherent control $\phi$ pulse [41], then the states are narrowly distributed around $m = \sin^2\left(\frac{\phi}{2}\right) N$ and we can approximate the effective coupling as:

$$g_{\text{eff}} = \frac{\sin(\phi)}{2} Ng. \tag{7}$$

The most important result here is that $g_{\text{eff}}$ scales like $N$ and not like $\sqrt{N}$ as expected from non-correlated interactions. This enchantment is due to the mutual quantum coherence between the emitters and is analogous yet complementary to the effect of free-electron–bound-electron resonant interaction [3]: There, each electron's wavefunction is bunched periodically so its constituents all see the same phase of a single oscillating emitter, whereas here, an ensemble of emitters are all organized so their oscillating phase is seen as constant by a single moving electron (independent of its wavefunction shape). Both effects give a resonant interaction between the emitters and the electrons, but from a complementary physical origin.

Another consequence of this complementarity is that whereas free electrons can induce Rabi oscillations on a single emitter in [3], in our work the ensemble of emitters induce Rabi oscillations on a single electron as can be shown through Eq. (6) [13-16]. Future works could investigate the interplay of these two types of Rabi oscillations when both the electron's and the emitter's mutual coherence are presented simultaneously.

There exist two types of superradiance: Dicke superradiance, where the emitters act as an effective big dipole, and free-electron superradiance, where the electrons act as an effective big charge (as used in klystrons [42], synchrotrons [43,44], and free-electron lasers (FELs) [45]). This work focuses on the first type, and thus our predictions apply even with just a single electron. Nevertheless, we note that these two types of superradiance can happen simultaneously, resulting in a double enhancement of EELS, and of the resulting cathodoluminescence signal [46]. The EELS cross-section is typically proportional to $(d \cdot e)^2$, with $d$ being the dipole strength and $e$ being the electron charge. Superradiance of the emitters and/or electrons enhance the cross-section by scaling $d \rightarrow N_a d$ and/or $e \rightarrow N_e e$ with $N_{a/e}$ being the number of emitters and electrons respectively.

## 3. Phase matching between a single free electron and quantum emitters

For an ensemble of excited emitters to satisfy a phase matched interaction with the electron, they need to be excited such that they remain inside the superradiant sub-Hilbert space following their (delayed) interactions with the electron. This implies that the relative phase between emitters separated by $\Delta z$ should be $\omega_0 \Delta z / v$. Since $v < c$, such a relative phase cannot be achieved by a free-space laser excitation, and so a material with an index of refraction $n$ can be utilized as illustrated in Fig. 1c. For a laser excitation propagating through such a material at angle $\theta$, the phase-matching condition translates to the following condition on the angle:

$$\cos \theta_C = \frac{c}{nv}. \tag{8}$$

This is exactly the Cherenkov condition [47,48] for phase matching between light and an electron. However, unlike all previous works in the field, here the phase matching is between the quantum emitters and an electron. All previous works that referred to superradiant effects in similar configurations (e.g., [49,50]) referred to the other type of superradiance – free-electron superradiance – arising from the number of free-electrons. Here, we access a different phenomenon – Dicke superradiance – arising from the quantum nature of the emitters and their correlated excitation.

To realize such an interaction, we imagine a prism such as in Fig. 1c, and multiple emitters such as quantum-dots [28,30] or perovskites [51-53] positioned on the surface of the prism. The laser excites the emitters with an angle matching the Cherenkov condition for an electron impinging at a grazing angle [21]. The EELS of an electron arriving *during* the laser excitation is expected to be dominated by the direct electron-laser interaction (enhanced by the same phase-matching condition [21]), rather than by the electron-emitters interaction. The signature of electron-emitters interaction should appear for an electron arriving after the laser excitation is over, and before the emitters lose the relative phase between them, quantified by the ensemble decoherence time $T_2^*$. Arrays of QDs can provide sufficient $T_2^*$: on the picoseconds scale even in room temperatures [54,55], which is long enough for typical interaction durations of electrons in grazing-angle configurations (hundreds of femtoseconds) [21]. Therefore, the necessary conditions to observe the electron-emitters phase-matched interaction exist using readily available materials and experimental configurations.

In Fig. 2, we look at the energy spectra of the post-interaction electron with multiple emitters after a $\pi/2$ pulse arriving at different angles. A $\pi/2$ pulse is considered to maximize the emitters' coherence (and hence, the effective coupling, see Eq. (7)). Fig. 2b shows that when the phase-matching condition is satisfied, the post-interaction EELS becomes much wider, indicating a stronger interaction. To quantify the enchantment, we define the effective coupling constant of one interaction according to:

$$g_{\text{eff}} \equiv \frac{\sigma_{\text{EELS}}}{\sqrt{2}\hbar\omega_0}, \tag{9}$$

where $\sigma_{\text{EELS}}$ is the standard deviation of the resulting energy distribution. This definition is consistent with the analogy to photo-induced nearfield electron microscopy in Eq. (6). In Fig. 2c, we plot the normalized effective coupling, confirming that the maximum coupling is achieved in the phase-matching condition, with a $\pi/2$ excitation.

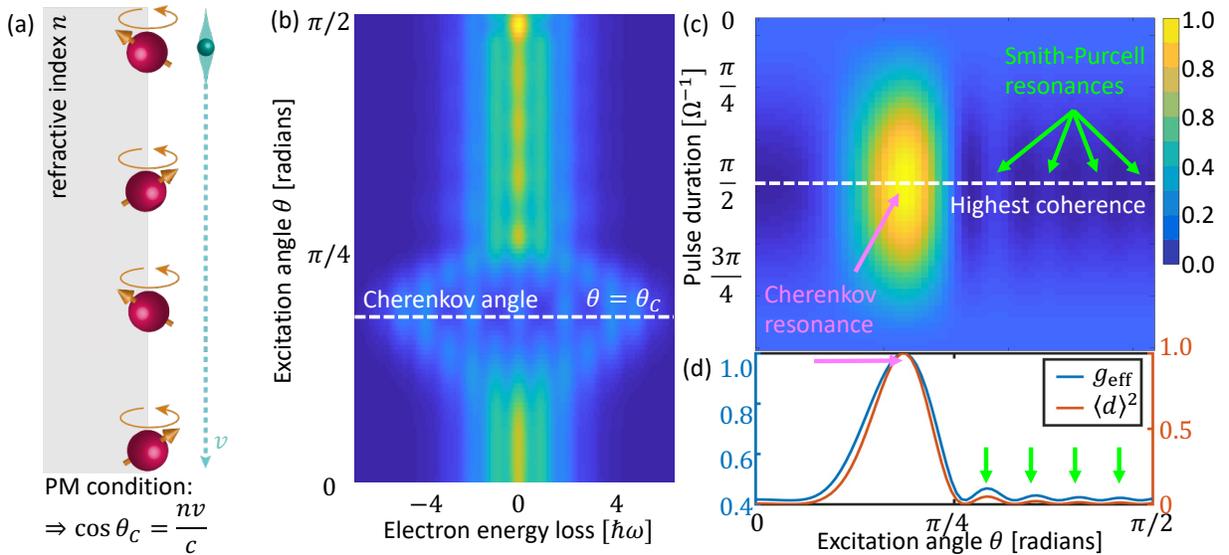

**Fig. 2. Phase-matched interaction between multiple emitters distributed over an extended area and a single electron that mediates their collective interaction. (a)** System schematic.

Emitters excited by a laser with its wave front tilted by an angle $\theta$, creating a relative delay between consequent emitters of $\Delta t = \cos\theta\, \Delta z\, n/c$. A free electron with velocity $v$ interacts with the emitters at a certain phase delay $\Delta t = \Delta z/v$, creating a resonant interaction when the time it takes to pass between consequent emitters matches the delay induced by the laser excitation. The condition for such a resonant interaction is Cherenkov-type ($\cos(\theta_C)\, n/c = 1/v$). This condition also enables the superradiant EELS phenomenon. **(b)** EELS as a function of excitation angle by a $\pi/2$ pulse. The strongest interaction is achieved at the Cherenkov angle $\theta_C$, satisfying phase matching between the electron and the emitters. **(c)** The effective coupling $g_{\text{eff}}$ as a function of the excitation pulse duration and angle. The Smith-Purcell-type resonances arise from the periodicity in the positions of the emitters (taken here to be 10nm, fitting a characteristic size of a quantum dot). If the emitters are placed in random locations, the Cherenkov-type resonance remains but the Smith-Purcell-type resonances vanish. **(d)** The effective coupling for a $\pi/2$ pulse, in comparison with the expectation value of the dipole as seen by the electron. All plots are normalized by peak value. The parameters for these plots are $v = 0.7c$, $N = 10$, $|g| = 0.5$.

Figs. 2c,d show oscillations in the effective coupling that we attribute to higher-order phase-matching arising from Smith-Purcell-like conditions (modified by the material's refractive index) due to the periodicity of the emitters' positions [56,57]. To understand this effect, we write the dipole expectation value as seen by the moving electron (see Fig. 2d):

$$|\langle d(\theta)\rangle|^2 = d_0^2 \left|\sum_i e^{i\omega_0 z_i\left(\frac{n\cos\theta}{c} - \frac{1}{v}\right)}\right|^2. \tag{10}$$

To find the resonant excitation angles marked in Figs. 2c,d, we require for the arguments of the exponents in Eq. (10) to be multiples of $2\pi$, which exactly yields a hybrid Cherenkov-Smith-Purcell condition as described in [58] for the inverse effect (i.e., for light emission by the emitters rather than for resonant electron interaction with pre-excited emitters). The periodicity in the locations of the emitters yields constructive interference in the interaction with the electron at these additional excitation angles besides the Cherenkov angle.

## 4. Superradiant ladder state reconstruction using free electrons

When the state of the emitters is part of the superradiant ladder, the entire population of the emitters is encoded on the free electron's post-interaction energy spectrum. This allows us to reconstruct the quantum state of the emitters using EELS. To show this, consider a general superradiant state of the form $|\psi_a\rangle = \sum c_m |m\rangle = \sum c_m S_+^m |gg \ldots g\rangle$ interacting with a free electron. The resulting EELS contains the probability of the electron to lose $k$ quanta of energy:

$$P_{-k} = \sum_m d_{km} |c_m|^2,$$

$$d_{km} = (\cos|g|)^N (i \tan|g|)^{k-m} \sqrt{m!\, k!\, (N-k)!\, (N-m)!} \sum_{n=0}^{m} \frac{(-1)^n (\tan|g|)^{2n}}{n!\, (m-n)!\, (k-m+n)!\, (N-k-n)!}. \quad (11)$$

Then, the population statistics can be constructed from the energy spectrum by inverting the matrix $d_{km}$. For a more general initial electron state $|\psi_e\rangle = \sum g_k |E_0 + k\hbar\omega_o\rangle$ that can be generated using coherent modulation of the free electron wavepacket [22,23], the final energy spectra will also include decodable information about the off-diagonal elements of the superradiant state (similar to [4]). For specific examples of how the state is encoded in the EELS, see Fig. 3.

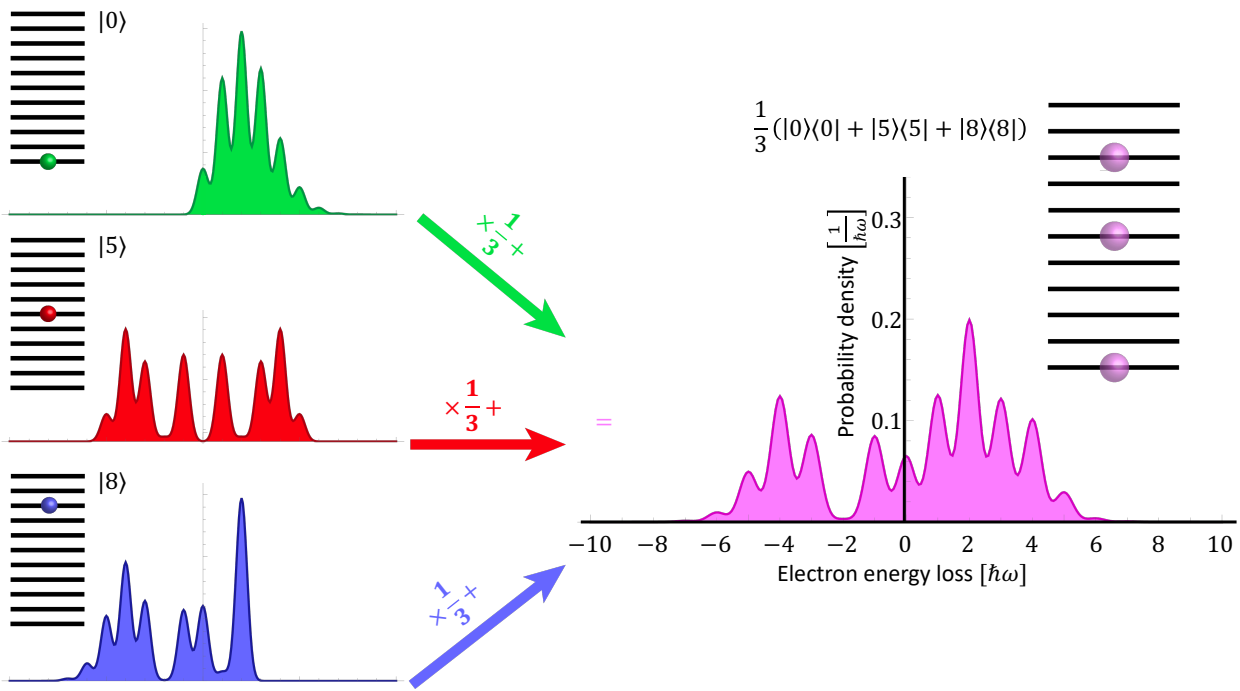

**Fig. 3. Extracting the population of the superradiant states from EELS measurements**. The EELS is found to be an incoherent weighted sum of the corresponding electron spectra of the different states on the ladder. Therefore, the population can be inferred by decomposing the spectrum according to Eq. (11). The parameters are taken to be the same as in Fig. 2.

## 5. Free-electron measurement of superradiant dynamics

The dynamics of Dicke Superradiance [31,32] includes a rapid relaxation of the identical emitters to their ground state, accompanied by emission of a short and intense pulse of light. This phenomenon results from the coherence and indistinguishability between the emitters, keeping them in the superradiant ladder during their relaxation dynamics. Many of the open questions about the process of Dicke superradiance relate to the population statistics of the emitters in the superradiant ladders, and the internal correlations between different ladder states. Typical measurements of superradiance rely on the emitted pulse intensity, which only reveals the mean-field dynamic of the emitters' population [32-34]. However, as shown by Bonifacio et al. [33,34], much interesting information lies beyond the mean-field approximation: for example, relating higher order correlations of the emitted light to correlations in the ladder states. We expect the full quantum state of superradiant light to depend on the correlations between the different superradiant states, and the photon statistics (and therefore the Mandel Q parameter [59]) to depends on the population statistics. Fig. 4a presents our proposal for extracting the population statistics of the emitters during superradiance using their interaction with a free electron.

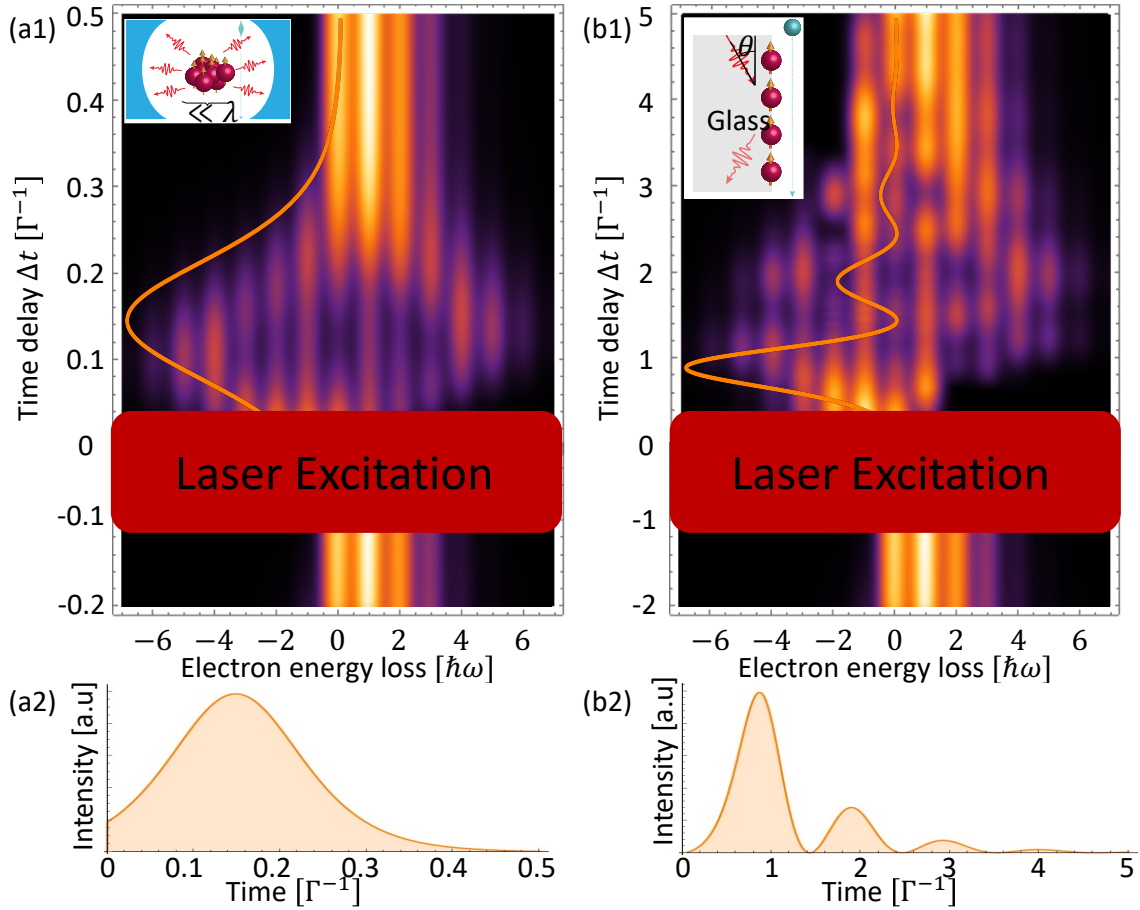

**Fig. 4. Extracting information about the superradiance dynamics from EELS**. The emitters are initiated at the fully excited states using a (π-pulse) laser excitation (marked on the panels). Panel **(a1)** describes a dense-ensemble situation as in Fig. 1a, and panel **(b1)** describes a grazing-angle configuration as in Fig. 1c. A free electron measures their state as a function of the time delay, showing the population relaxation during the superradiant emission. The EELS at each time delay enables extracting the emitters' population on the superradiant ladder according to Eq. (11). Panels (a2) and (b2) present the corresponding time-dependent intensity plots. The parameters for these plots are $v = 0.7c$, $N = 30, |g| = 0.2$.

Fig. 4 exemplifies EELS measurements for extracting the transient population statistics of the emitters. The intensity envelope is encoded on the electron's energy spectrum, together with the entire population statistics. The emitters are excited using a laser pulse at time $t = 0$ and interact with an electron at a later time $t = \Delta t$. Using the scheme presented in section 4, the emitters' state can be reconstructed (for times shorter than decoherence $T_2^*$ so their state remains in the sub-Hilbert space). This experiment can be repeated for different time delays to reconstruct the population dynamic.

Fig. 4b exemplifies this scheme for long-sample superradiance [32], using the truncated-Wigner approximation [60] to calculate the superradiant dynamic (see SM section 3). The phase-matching condition on the excitation laser provides the emitters with a phase relation enabling superradiance and decay inside the superradiant ladder. The long-sample nature gives rise to a characteristic fluctuating intensity pattern. To the best of our knowledge, solving the population dynamic in such a system beyond the mean-field approximation is a difficult problem without a known analytical solution [32,35]; this motivates an experimental investigation of the scheme we present here.

## 6. Discussion and outlook

Looking at the grater field of EELS gives our theory a wider context: as an instance of **EELS for out-of-equilibrium systems**, showing the importance of mutual coherence in matter for interactions with free electrons. This goes beyond the conventional theory of EELS [7] that rely on systems in equilibrium and on linear response theory [61-63]. Our work thus contributes to our ongoing efforts toward a general understanding of out-of-equilibrium EELS, which was so far only investigated for electron interaction with a single quantum emitter [4]. Advances in ultrafast electron microscopy [12-21,38] now make it possible to investigate this new area experimentally.

So far, the coherence of quantum emitters was seen as important only for interactions with specially shaped electrons [3-6]. Our work shows that when the coherence involves multiple systems, it affects the interaction even with unshaped electrons, which opens new paths for experimental observations of such phenomena. Furthermore, coherent interaction of free electrons with bound electron systems [3-7,11] is yet to be observed experimentally due to the intrinsically weak coupling. Our proposal of utilizing an ensemble of identical emitters interacting with a single electron can enhance the interaction significantly, enabling experiments in new complementary systems.

The Dicke superradiance process is typically probed optically by measuring the emitted pulse. However, understanding the superradiant dynamic on small scales, at spatial resolutions beyond the reach of optical methods, remains a great challenge. The electron interaction opens the possibility of using electron pulses to image the dynamics of Dicke superradiance, extracting the intermediate populations on the superradiant ladder with simultaneous high spatial and temporal resolution, as in ultrafast transmission electron microscopy [12,38].

To conclude, our work investigated for the first time the coherent interaction between a free electron and an ensemble of emitters. We found resonances in the electron-emitters interaction, facilitated by the mutual coherence between the emitters. Free electrons can be utilized to investigate superradiance over spatial and temporal resolutions that are inaccessible to conventional optical techniques. Besides the enhanced resolution, the electrons provide quantum information regarding the emitters' state, which is inaccessible using conventional measurement techniques in superradiance. Further information of the quantum state of the emitters, such as their off-diagonal coherences [4,5] can be accessed as well by pre-shaping the electron before its interaction [3-5], potentially enabling even stronger superradiant-EELS resonances.